\documentclass{ws-procs9x6-cpt19}
\usepackage{accents}
\usepackage{color}
\usepackage{xcolor}

\def\lsim{\mathrel{\rlap{\lower4pt\hbox{\hskip1pt$\sim$}}
    \raise1pt\hbox{$<$}}}
\def\gsim{\mathrel{\rlap{\lower4pt\hbox{\hskip1pt$\sim$}}
    \raise1pt\hbox{$>$}}}

\newcommand{\beq}{\begin{eqnarray}}
\newcommand{\eeq}{\end{eqnarray}}

\begin{document}

\title{Highlights from the 7 year High Energy Starting Event sample in Icecube}

\author{
  Kareem Farrag$^{1,2}$
%  Carlos~A.~Arg\"{u}elles$^3$, 
%  Teppei Katori$^1$, and
%  Shivesh Mandalia$^1$
}

\address{
  $^1$Queen Mary University of London, E1 4NS, UK\\
  $^2$University of Southampton, Southampton, SO17 1BJ, UK\\
  %$^3$Massachusetts Institute of Technology, Cambridge, MA 02139, USA\\
}

\author{On behalf of the IceCube Collaboration}

\begin{abstract}
Here we outline the main highlights from the 7 year High Energy Starting Events (HESE) event sample. The next new physics search using astrophysical neutrino flavor data is described, where we reach the Planck scale for the first time.
\end{abstract}
\bodymatter
%%%%%%%%%%%%%%%%%%
%  HESE Intro    %
%%%%%%%%%%%%%%%%%%
\section{High Evergy Starting Events (HESE) in Icecube}\label{sec:hese}
The HESE data sample in IceCube~\cite{Aartsen:2016nxy} is selected to extract a sample of astrophysical neutrinos with high purity~\cite{Aartsen:2013jdh,Stachurska:2019srh}. Events are accepted into the sample if the interaction vertex is contained in a subvolume of IceCube, defined by the inner part of the detector as the outer most layers are used as a veto region to reject atmospheric backgrounds~\cite{Arguelles:2018awr}. We find 102 events observed over 2635 days with 60 events above 60 TeV in deposited energy. HESE is a low atmospheric background event selection used in IceCube to study astrophysical neutrinos, including dark matter (DM) searches and anomalous spacetime effects through the astrophysical neutrino flavour composition.

\begin{figure}[!ht]\label{fig:hese}
\begin{minipage}{0.49\textwidth}
    {\includegraphics[scale=0.23 ]{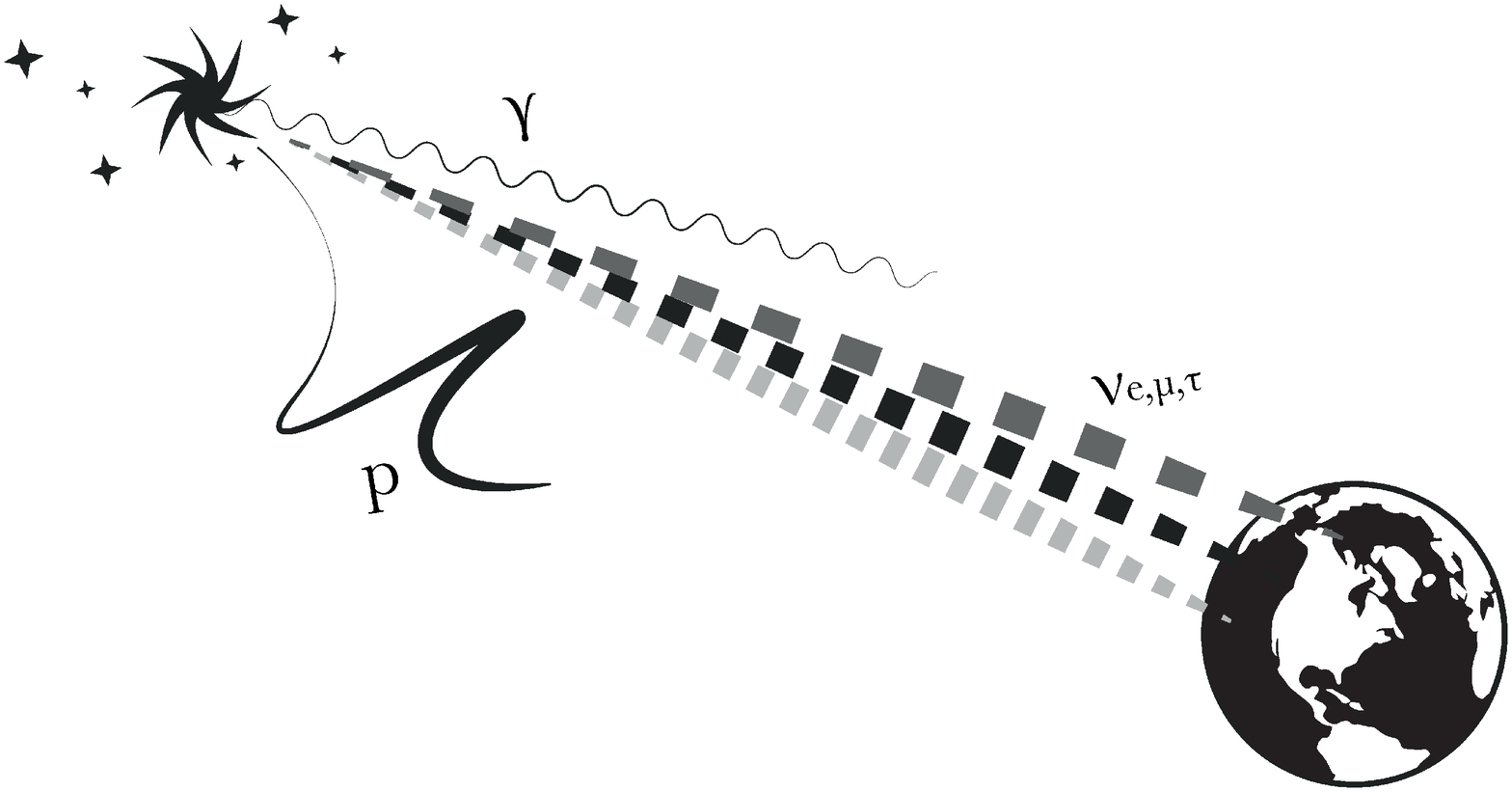}}
    \end{minipage}\hfill
     \begin{minipage}{0.49\textwidth}
    {\includegraphics[scale=0.25 ]{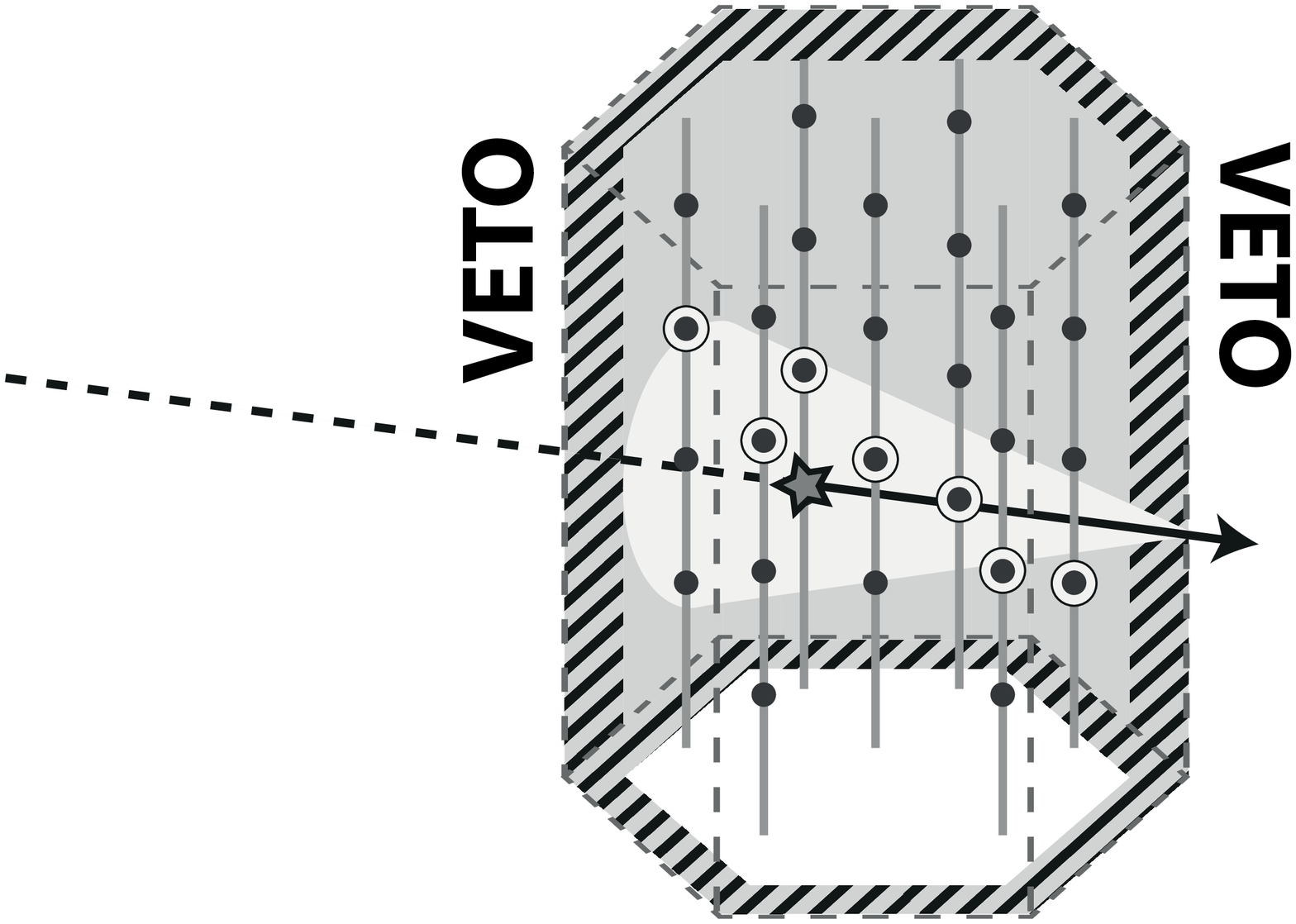}}
     \end{minipage}\hfill
    \caption{Left shows illustration of astrophysical neutrino interferometry. Neutrinos at source and production exist in flavor eigenstates. However, they travel in the so-called propagation basis. This basis evolves with time according to their effective Hamiltonian. Right illustration depicts how the HESE selection is done. Trigger demands that the interaction vertex be fully contained (bound by the dashed region). Events require more than 6000 photoelectrons to ensure (to 99.999\%) that cosmic ray muons would produce enough light in the veto region to be excluded~\cite{Aartsen:2013jdh}.}
\end{figure}

\begin{figure}[!ht]\label{fig:powerspectrum}
\begin{minipage}{0.50\textwidth}
 \includegraphics[width=\textwidth]{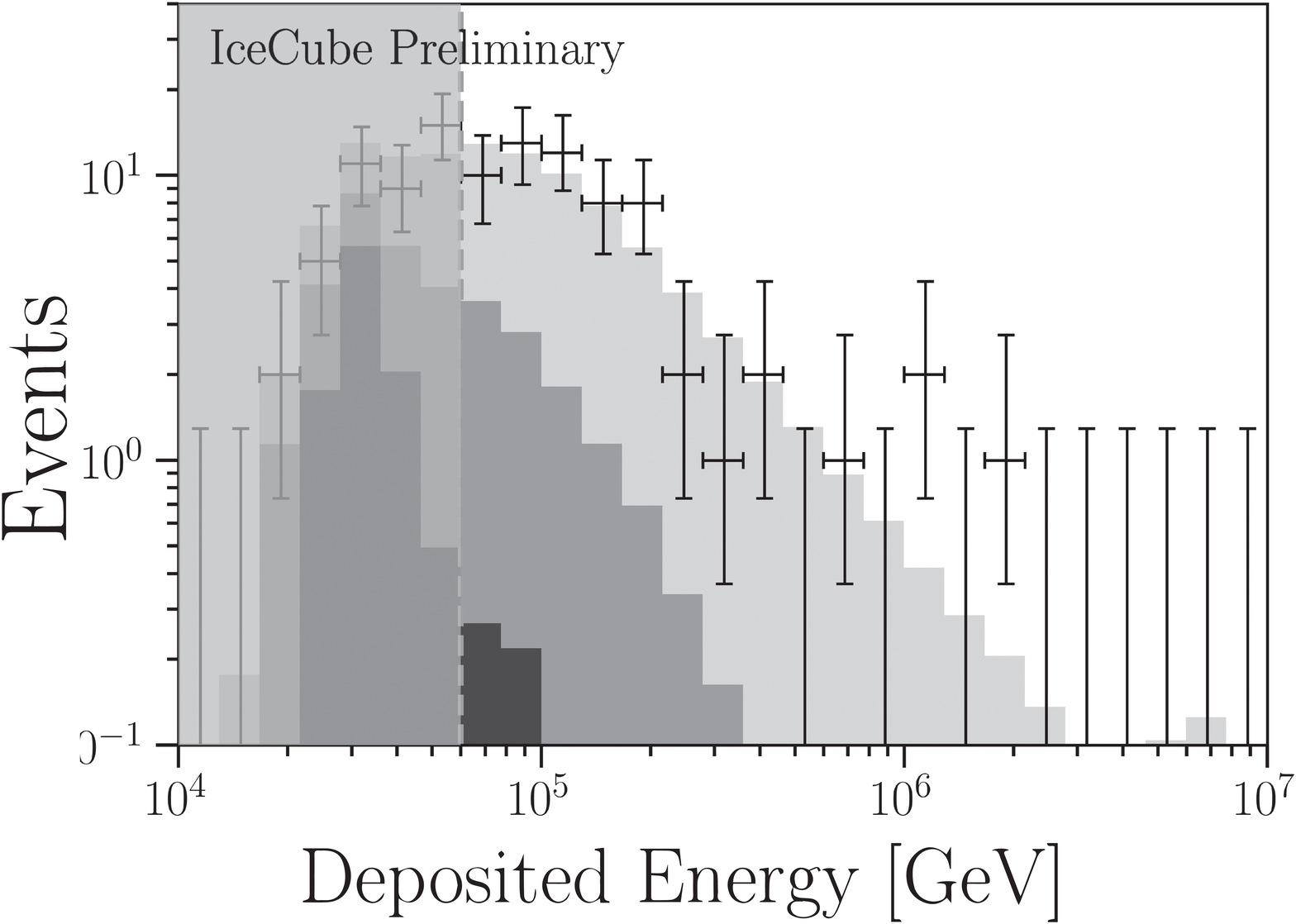}
\end{minipage}\hfill 
\begin{minipage}{0.50\textwidth}
  \includegraphics[width=\textwidth]{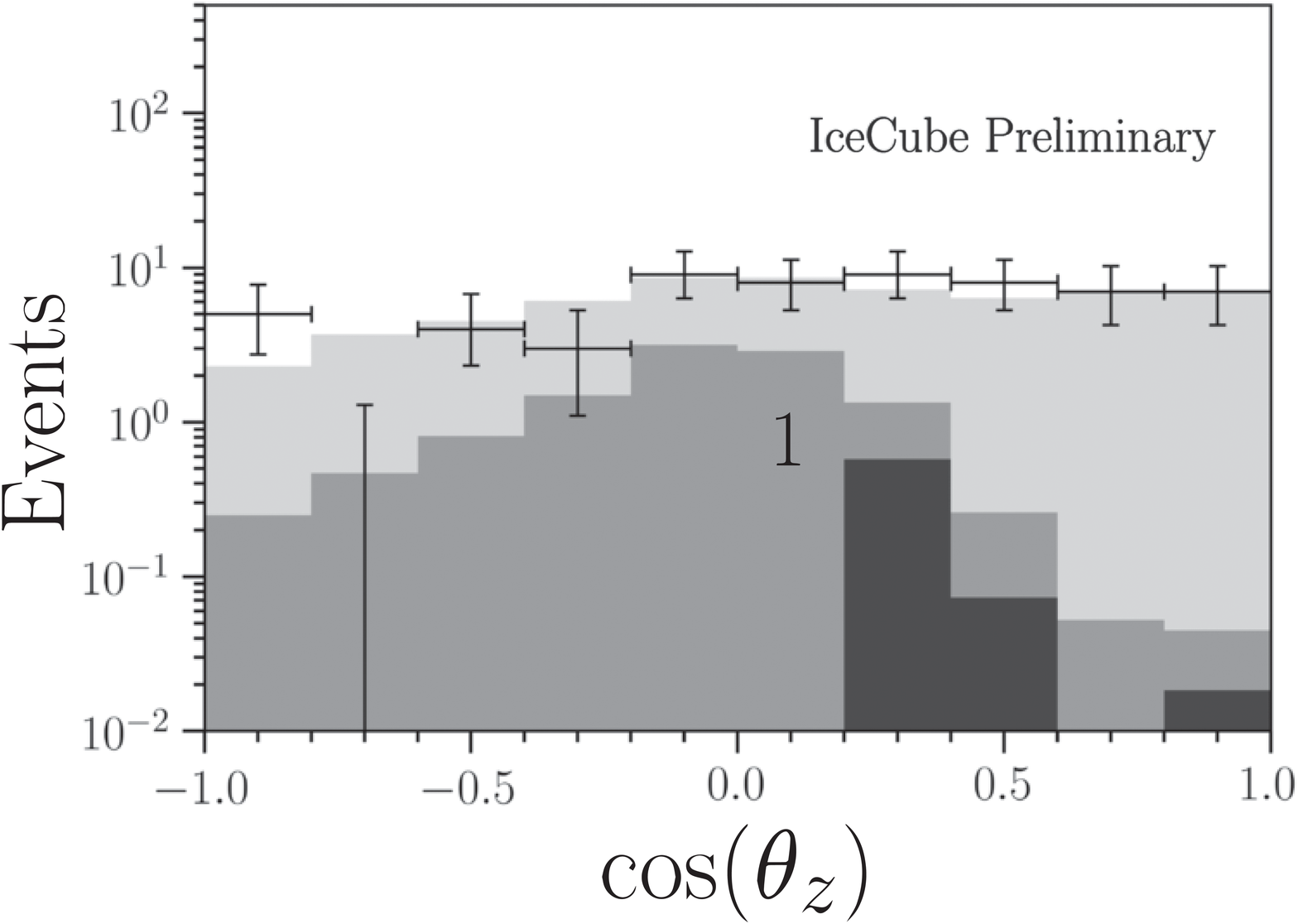}
\end{minipage}

\vspace{5pt}
\fbox{\begin{minipage}{0.97\textwidth}
 {\includegraphics[scale=0.025]{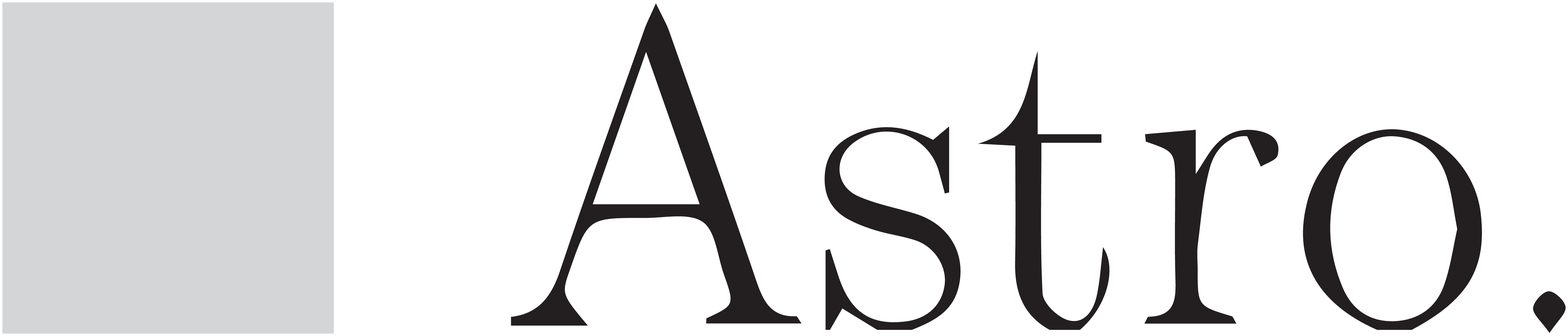}}\hfill
 {\includegraphics[scale=0.025]{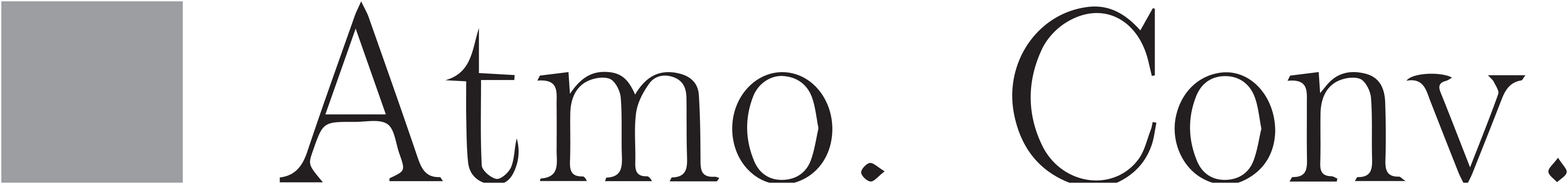}}\hfill
 {\includegraphics[scale=0.025]{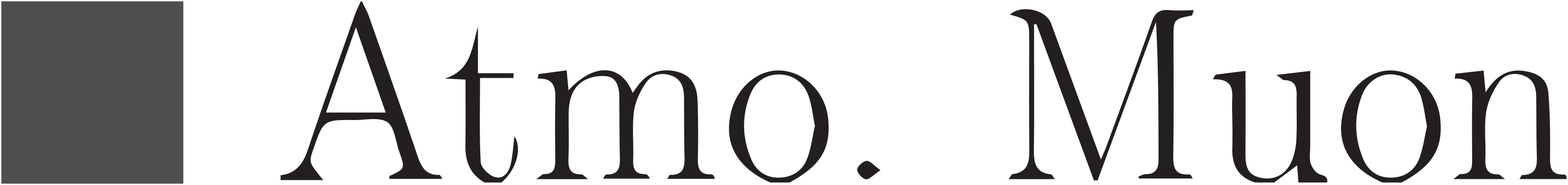}}\hfill
 {\includegraphics[scale=0.025]{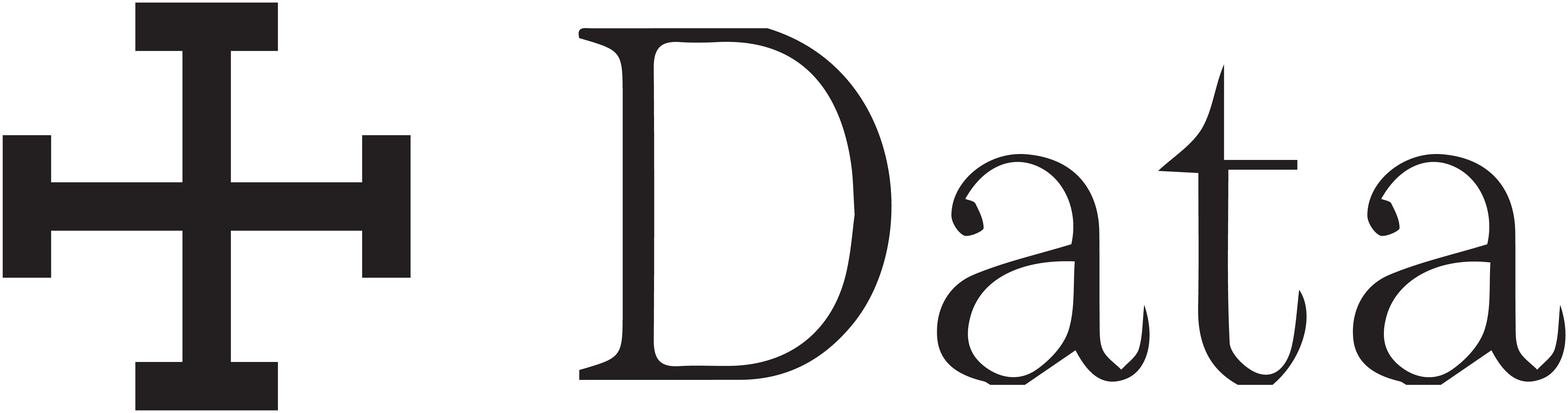}}
\end{minipage}
 }
    \caption{Left plot shows HESE 7 energy distribution. The $x$-axis is deposited energy estimate by a neutrino event interaction. Atmospheric conventional and muon estimates in the sample are mid and dark gray regions respectively. The astrophysical component in light gray has a harder energy spectrum with spectral index $\gamma\sim 2.9$. The number of events in each bin are marked by a cross. Right plot shows zenith distribution for the HESE sample.}
\end{figure}

%$%%%%%%%%%%%%%%%%%
% RECONSTRUCTION  %
%%%$%%%%%%%%%%%%%%%
Key changes in the event reconstruction include global changes to the ice model. This includes both ice anistropy and tilt effects. HESE 7 is consistent with a single power law fit with spectral index of $\gamma \sim 2.9$. Systematic uncertainties include contributions from atmospheric neutrino fluxes as well as detector systematics. We show the energy and angular distributions in Fig.~\ref{fig:powerspectrum}. 
%%%%%%%%%%%
% Flavour %
%%%%%%%%%%%
\begin{figure}[!ht]\label{fig:astro}
    \begin{minipage}{\textwidth}
    \centering
    \includegraphics[scale=0.23]{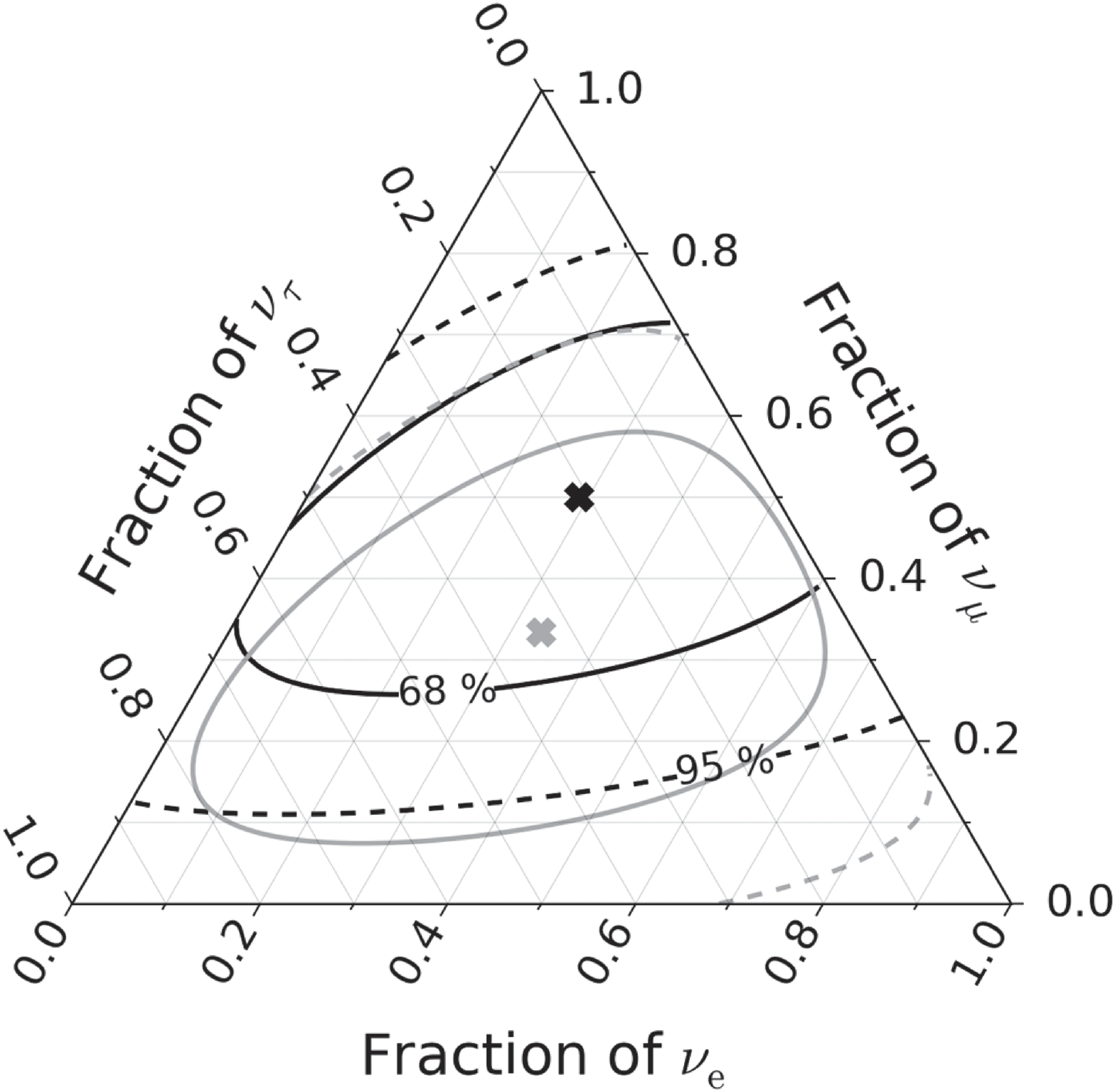}
    \end{minipage}
    \begin{minipage}{\textwidth}
    \centering
    \fbox{
    $\text{IceCube Work In Progress}$
    }
    \end{minipage}
    \caption{Ternary plot shows current sensitivity and data contour, which are consistent with $(1:1:1)$ flavor ratio on Earth. Gray contour shows sensitivity with best fit crossed at $(1:1:1)$. Black contour shows data contour for HESE with ternary ID topology with cross at best fit $(0.29:0.50:0.21)$. Solid and dashed lines show 68\% and 95\% credibility regions respectively.}
\end{figure}
The preliminary best fit for the flavour composition of diffuse neutrinos is $(0.29:0.50:0.21)$, where the current flavor contour is consistent with $(1:1:1)$. Note zero tau events cannot be ruled out with best fit $E^{-2.9}$ spectrum. Contours are computed with Wilk's theorem, and work in this avenue is ongoing. 
A BSM search using the Standard Model Extension~\cite{Kostelecky:2003fs} was performed, attempting to detect anomalous flavor ratios. Such ratios could arise due to the presence of effective operators at high energy scales. Three source flavour compositions of the form ($f_e: f_\mu: f_\tau$) are believed to dominate, namely  $(1:2:0)$, $(1:0:0)$ and $(0:1:0)$. We expect to place the most stringent limits on higher order new physics operators.
%dimension 6 operator around $10^{-45}~\textup{GeV}^{-2}$.
Furthermore, two tau candidate neutrinos have been observed in HESE, with double cascade energies $E\sim 100~\textup{TeV}$ and $1.8~\textup{PeV}$. This corresponds to the first astrophysical tau neutrino candidates in IceCube. Current work is ongoing to quantify their significance. 
%%%%%%%%%%%%%%%%%%%%%
%   Conclusion      %
%%%%%%%%%%%%%%%%%%%%%
In conclusion, our first ever search of Lorentz violation (LV) was done using astrophysical neutrinos. Most notably, our limit resides several orders of magnitude below the Planck scale. More work is needed to constrain all source flavor paradigms. Our technique reaches the quantum gravity regime for the first time. Future searches will be able to probe the region of flavor space currently inaccessible with 7 years of HESE data. 


\begin{thebibliography}{}
% Reference on IceCube
%\cite{Aartsen:2016nxy}
\bibitem{Aartsen:2016nxy} 
  M.~G.~Aartsen {\it et al.} [IceCube Collaboration],
  %``The IceCube Neutrino Observatory: Instrumentation and Online Systems,''
  JINST {\bf 12}, no. 03, P03012 (2017)
%   doi:10.1088/1748-0221/12/03/P03012
%   [arXiv:1612.05093 [astro-ph.IM]].
  %%CITATION = doi:10.1088/1748-0221/12/03/P03012;%%
  %158 citations counted in INSPIRE as of 13 Jun 2019
  
  %HESE Reference
  %\cite{Aartsen:2013jdh}
\bibitem{Aartsen:2013jdh} 
  M.~G.~Aartsen {\it et al.} [IceCube Collaboration],
  %``Evidence for High-Energy Extraterrestrial Neutrinos at the IceCube Detector,''
  Science {\bf 342}, 1242856 (2013)
%   doi:10.1126/science.1242856
%   [arXiv:1311.5238 [astro-ph.HE]].
  %%CITATION = doi:10.1126/science.1242856;%%
  %882 citations counted in INSPIRE as of 13 Jun 2019

  %Juliana's proceeding reference
%\cite{Stachurska:2019srh}
\bibitem{Stachurska:2019srh} 
  J.~Stachurska [IceCube Collaboration],
  %``IceCube High Energy Starting Events at 7 Years -- New Measurements of Flux and Flavor,''
  EPJ Web Conf.\  {\bf 207}, 02005 (2019)
%   doi:10.1051/epjconf/201920702005
%   [arXiv:1905.04237 [hep-ex]].
  %%CITATION = doi:10.1051/epjconf/201920702005;%%
  
\bibitem{Arguelles:2018awr} 
  C.~A.~Arg{\"u}elles {\it et al.}, 
  %S.~Palomares-Ruiz, A.~Schneider, L.~Wille and T.~Yuan,
  %``Unified atmospheric neutrino passing fractions for large-scale neutrino telescopes,''
  JCAP {\bf 1807}, no. 07, 047 (2018).
  %doi:10.1088/1475-7516/2018/07/047
  %[arXiv:1805.11003 [hep-ph]].
  %%CITATION = doi:10.1088/1475-7516/2018/07/047;%%
  %2 citations counted in INSPIRE as of 13 Jun 2019

%\cite{Kostelecky:2003fs}
\bibitem{Kostelecky:2003fs} 
  V.~A.~Kostelecky,
  %``Gravity, Lorentz violation, and the standard model,''
  Phys.\ Rev.\ D {\bf 69}, 105009 (2004)
%   doi:10.1103/PhysRevD.69.105009
%   [hep-th/0312310].
  %%CITATION = doi:10.1103/PhysRevD.69.105009;%%
  %934 citations counted in INSPIRE as of 13 Jun 2019

  
\end{thebibliography}
\end{document}